\documentclass[twocolumn]{IEEEtran}
\usepackage{graphicx} 
\usepackage{url} 
\usepackage{tcolorbox}
\usepackage{hyperref}
\usepackage{multirow}
\usepackage{geometry}
\usepackage{multirow}
\usepackage{makecell}
\usepackage{booktabs}      
\usepackage{adjustbox}     
\usepackage{caption}       
\usepackage{url}

\title{Harnessing Multi-Agent LLMs for Complex Engineering Problem-Solving: A Framework for Senior Design Projects}

\author{
    Abdullah Mushtaq\IEEEauthorrefmark{1}, 
    Muhammad Rafay Naeem\IEEEauthorrefmark{1},  Ibrahim Ghaznavi\IEEEauthorrefmark{1},
    Muhammad Imran Taj\IEEEauthorrefmark{2},  
    Imran Hashmi\IEEEauthorrefmark{3},  
    Junaid Qadir\IEEEauthorrefmark{4} \\

    \IEEEauthorblockA{\IEEEauthorrefmark{1}Department of Computer Science, Information Technology University, Lahore, Pakistan \\
    \{bscs20078, bscs20004, ibrahim.ghaznavi\}@itu.edu.pk} \\

    \IEEEauthorblockA{\IEEEauthorrefmark{2}College of Interdisciplinary Studies, Zayed University, Dubai, UAE \\
    MuhammadImran.Taj@zu.ac.ae} \\

    \IEEEauthorblockA{\IEEEauthorrefmark{3}Department of Computer Science, University of Oxford, Oxford, UK \\
  imran.hashmi@cs.ox.ac.uk\\
}
\IEEEauthorblockA{\IEEEauthorrefmark{4}Department of Computer Science and Engineering, Qatar University, Doha, Qatar \\
    jqadir@qu.edu.qa}

}

\newtcolorbox[auto counter, list inside=pabox]{pabox}[2][]{%
colframe=green!70!black,
colback=green!5!white,
fonttitle=\bfseries,
coltitle=white,
boxrule=0.75pt,       
arc=6pt,              
outer arc=6pt,
width=\columnwidth,   
left=6pt,             
right=6pt,            
top=6pt,              
bottom=6pt,            
before skip=8pt,    
after skip=8pt,     
title=Box~\thetcbcounter: #2,#1}

\begin{document}

\maketitle


\begin{abstract}
Multi-Agent Large Language Models (LLMs) are gaining significant attention for their ability to harness collective intelligence in complex problem-solving, decision-making, and planning tasks. This aligns with the concept of the wisdom of crowds, where diverse agents contribute collectively to generating effective solutions, making it particularly suitable for educational settings. Senior design projects, also known as capstone or final year projects, are pivotal in engineering education as they integrate theoretical knowledge with practical application, fostering critical thinking, teamwork, and real-world problem-solving skills. In this paper, we explore the use of Multi-Agent LLMs in supporting these senior design projects undertaken by engineering students, which often involve multidisciplinary considerations and conflicting objectives, such as optimizing technical performance while addressing ethical, social, and environmental concerns. We propose a framework where distinct LLM agents represent different expert perspectives, such as problem formulation agents, system complexity agents, societal and ethical agents, or project managers, thus facilitating a holistic problem-solving approach. This implementation leverages standard multi-agent system (MAS) concepts such as coordination, cooperation, and negotiation, incorporating prompt engineering to develop diverse personas for each agent. These agents engage in rich, collaborative dialogues to simulate human engineering teams, guided by principles from swarm AI to efficiently balance individual contributions towards a unified solution. We adapt these techniques to create a collaboration structure for LLM agents, encouraging interdisciplinary reasoning and negotiation similar to real-world senior design projects. To assess the efficacy of this framework, we collected six proposals of engineering and computer science of typical senior capstone projects and evaluated the performance of Multi-Agent and single-agent LLMs using both custom-designed metrics developed in consultation with engineering faculty and some widely used NLP-based metrics. These metrics cover technical quality, ethical considerations, social impact, and feasibility, ensuring that our evaluation aligns with the educational objectives of engineering design. Our findings suggest that Multi-Agent LLMs can provide a richer, more inclusive problem-solving environment compared to single-agent systems, offering a promising tool for enhancing the educational experience of engineering and computer science students by simulating the complexity and collaboration of real-world engineering and computer science practice. By supporting senior design projects, this tool not only aids in achieving academic excellence but also prepares students for the multifaceted challenges they will face in their professional engineering careers.
\end{abstract}

\begin{IEEEkeywords}
Large Language Models, Gen AI, LLM Agents, LLM-Based Multi-Agent Systems, Multi-Agent Collaboration, Agentic AI, Autonomous LLM Agents, LLM in Engineering Applications.
\end{IEEEkeywords}

\section{Introduction}
The senior design project (SDP), also known as the capstone or final year project, is a vital component of engineering and computer science education \cite{fernando2022work}. It offers students an opportunity to apply their theoretical knowledge in tackling complex, real-world engineering problems, providing an authentic learning experience that integrates the essential competencies required for 21st-century engineers \cite{qadir2020engineering, qadir2020student}. Accrediting bodies, such as the US Accreditation Board for Engineering and Technology (ABET) emphasize the importance of SDPs as key opportunities for students to engage with the kinds of complex, multidisciplinary challenges they will encounter in professional practice. The problems addressed in SDPs typically involve no straightforward solutions; instead, they require students to balance competing objectives, such as optimizing technical performance while addressing ethical, environmental, and social concerns.

A hallmark of modern engineering practice is the increasingly globalized context in which engineers operate. Engineering products must be designed to meet the needs of a global audience, often requiring the integration of diverse perspectives. This necessitates cultural intelligence---the ability to navigate different cultural contexts and work effectively with stakeholders from various backgrounds. Generative AI, particularly LLMs, offers an innovative way to simulate these diverse perspectives, providing students with an environment to engage in interdisciplinary problem-solving \cite{qadir2023engineering, johri2023generative}. These LLM agents represent different expert viewpoints and foster a collaborative approach that reflects real-world engineering practices.

The ability to solve complex problems is an essential competence in both education and professional life, as individuals increasingly face challenges arising from globalization and digitalization. A complex problem occurs when a person seeks to achieve a goal for which no clear or straightforward solution is available. Unlike non-complex problems, complex problem-solving (CPS) involves dynamic and opaque barriers, where the initial information is incomplete or subject to change. This definition, as articulated by Fischer, Greiff, and Funke, emphasizes that complex problems require adaptive thinking and flexibility in problem-solving approaches \cite{fischer2012process, frensch1995framework}. In the context of engineering, the complexity is further amplified by the need to balance multiple, often conflicting objectives. Systems thinking, as introduced by Peter Senge in The Fifth Discipline, highlights the necessity of understanding the interconnections within complex systems, encouraging engineers to consider multiple perspectives and the broader implications of their solutions \cite{senge1990fifth}.

By leveraging the principle of the ``wisdom of crowds''---the idea that large groups of diverse, independent individuals can collectively arrive at better decisions, solutions, and predictions than any single expert, as outlined by Surowiecki in The Wisdom of Crowds \cite{surowiecki2004wisdom}---MAS has the potential to enable collective intelligence through emergent behavior. While intelligence is not inherent in MAS by design, structured interactions, and coordination mechanisms among agents allow complex and intelligent behaviors to develop collectively, leading to the emergence of intelligence \cite{wooldridge2009introduction}. The framework we propose builds on this idea, using LLM agents to simulate the interactions between diverse expert perspectives, such as problem formulation agents, systems complexity agents, and ethical and societal agents. In this way, LLM agents in the proposed MAS can help students develop the critical thinking and collaboration skills they need to succeed in globalized, multidisciplinary engineering environments.

The contributions of this paper are both pedagogical and technical. From a pedagogical perspective, this work introduces a novel framework for assisting supervisors and training engineering students in complex problem-solving using multi-agent LLMs within the context of SDPs. This approach leverages the capabilities of diverse LLM agents that represent different expert perspectives, simulating real-world interdisciplinary collaboration and enhancing critical thinking. Technically, the paper advances the state of the art by developing a framework that integrates MAS with LLMs to simulate multi-faceted engineering scenarios. This approach makes our framework capable of more effective technical problem-solving while incorporating ethical, social, and environmental dimensions into the decision-making process, offering a holistic solution to training engineers in globalized and complex environments.

The remainder of this paper is organized as follows. Section \ref{sec:background} discusses background and related Work, introducing foundational concepts in Agent-Based Modeling and multi-agent systems, and highlighting previous efforts in applying AI tools for complex problem-solving in engineering education. Section \ref{sec:methodology} outlines the Methodology, detailing the use of Multi-Agent LLMs to support SDPs and the evaluation criteria. In Section \ref{sec:results}, we present the results, followed by a discussion in Section \ref{sec:discussions}, reflecting on the pedagogical and technical implications. Finally, Section \ref{sec:conclusions} concludes the paper, summarizing contributions and suggesting future research directions.


\section{Background and Related Work}
\label{sec:background}

\subsection{The Role of SDPs in Engineering Education}

SDPs, also referred to as capstone or final-year projects, are a crucial component of computing and engineering education. As required by ABET and other accrediting bodies, SDPs serve as a culminating experience that integrates the knowledge and skills students have acquired over their academic careers. These projects require students to solve complex, real-world engineering problems that involve conflicting objectives, trade-offs, and ethical considerations. For instance, engineers must often balance cost efficiency with environmental sustainability, or optimize technical performance while adhering to regulatory constraints.

In these settings, teamwork and collaboration across diverse skills and perspectives are essential. Given the globalized nature of modern engineering practice, students are also required to consider cultural and social factors when designing engineering solutions. This is especially important as engineers increasingly work in global teams and must develop products that meet the needs of an international audience. Therefore, engaging with multiple stakeholders and considering the broader societal and cultural contexts of engineering solutions are vital competencies that SDPs aim to cultivate.

\subsection{Complex Problem Solving and Diversity Dividend}

Complex engineering problems, as per the definition of ABET\footnote{\url{https://www.abet.org/accreditation/accreditation-criteria/criteria-for-accrediting-engineering-programs-2022-2023/}}, have the following attributes, ``involving wide-ranging or conflicting technical issues, having no obvious solution, addressing problems not encompassed by current standards and codes, involving diverse groups of stakeholders, including many component parts or sub-problems, involving multiple disciplines, or having significant consequences in a range of contexts.''

Several frameworks have been proposed that leverage diversity to foster effective problem-solving. One well-known approach is Edward de Bono's Six Thinking Hats framework, which encourages individuals to look at problems from multiple perspectives---logical, emotional, and creative, among others. This structured, yet flexible, approach is particularly effective when integrated with MAS, as it mirrors the interdisciplinary thinking necessary to solve complex engineering problems. Scott Page's book \textit{The Difference} \cite{page2007diversity} emphasizes the value of diversity in problem-solving, particularly in complex systems. Page argues that teams composed of individuals with different skills, experiences, and perspectives are better equipped to solve complex problems than homogenous teams. This aligns with the wisdom of crowds principle, which suggests that collective intelligence can outperform individual expertise, particularly when the group is diverse and composed of independent thinkers. Similarly, Minsky's \textit{The Society of Mind} \cite{minsky1986society} underscores the importance of diverse cognitive processes in problem-solving, positing that complex thought emerges from the interaction of simpler, specialized processes.

\subsection{Multi-Agent Systems and Agent-Based Modeling}

MAS is widely applied in fields like robotics, artificial intelligence, and complex systems modeling to simulate autonomous agents acting within dynamic environments. While each agent operates independently, their collective behavior can lead to emergent properties—patterns or outcomes that are challenging to predict solely from individual actions. In engineering education, MAS can be used to simulate stakeholder interactions in complex projects, enabling students to experience realistic decision-making, collaboration, and problem-solving scenarios.

Agent-Based Modeling (ABM) extends MAS by offering a detailed computational framework for simulating interactions between agents and their environments. ABM is especially valuable for examining systems where individual behaviors directly influence collective outcomes, making it ideal for complex social or engineering contexts. In SDPs, ABM can simulate interactions among engineers, project managers, regulators, and other stakeholders, giving students practical insight into the collaborative and often unpredictable nature of professional practice \cite{wang2024survey, guo2024large}. These simulations provide a flexible and dynamic approach for modeling complex systems, allowing students to explore the consequences of various decisions within a controlled environment

\begin{figure*}[htbp]
    \centering
    \includegraphics[width=\linewidth]{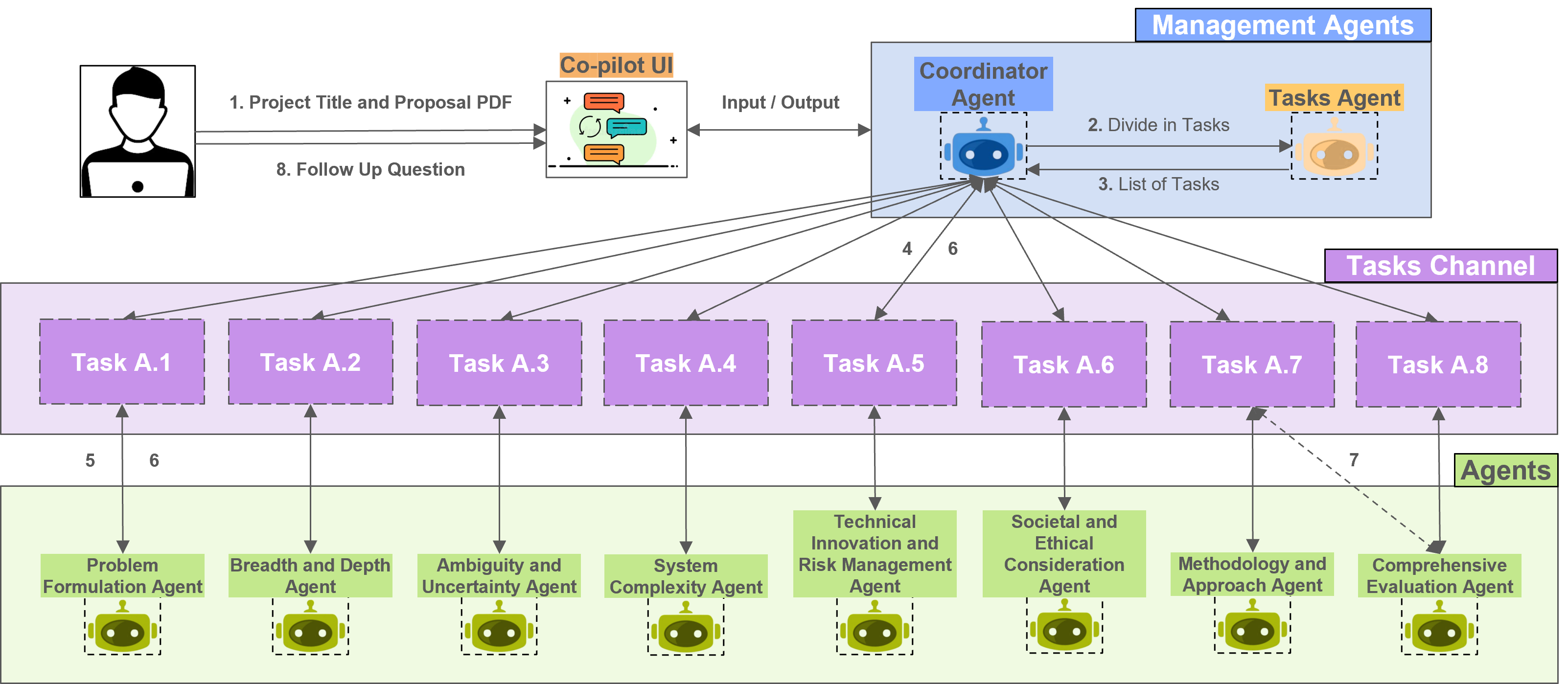}
   \caption{\textbf{Workflow Overview}: 
\textit{1) Input Submission:} The student enters the project title and proposal PDF.
\textit{2) Proposal Processing:} Coordinator Agent forwards details to Tasks Agent to generate focused tasks.
\textit{3) Task Generation:} Tasks Agent returns a list of tasks.
\textit{4) Task Distribution:} Tasks are sent to the Tasks Channel.
\textit{5) Task Assignment:} Tasks are assigned to relevant agents.
\textit{6) Output Generation:} Agents produce outputs and send them to the Tasks Channel and Coordinator Agent.
\textit{7) Input Linking:} Outputs from one agent can serve as inputs for others if needed.
\textit{8) User Output:} Final results are displayed in the interface, with summaries and detailed analysis.}
    \label{fig:System_Design}
\end{figure*}

\subsection{MAS and LLMs in Educational Contexts}

The integration of MAS and LLMs offers significant promise for enhancing engineering education, especially in tackling complex, real-world problems. By simulating interactions between diverse autonomous agents, MAS allows educators and students to explore the trade-offs, dilemmas, and decision-making processes typical in multidisciplinary projects. These systems mirror real-world engineering dynamics, where professionals must collaborate across domains, negotiate conflicting objectives, and optimize under constraints.

LLMs further enhance this framework by representing expert personas---such as problem formulation agents, project managers, and ethical and societal agents---enabling students to engage in interdisciplinary dialogue. Studies have shown that this approach leads to more innovative and robust solutions by leveraging the ``wisdom of crowds'' effect, as described by Surowiecki \cite{surowiecki2004wisdom}. By fostering diverse perspectives, students gain deeper insights into the complexities of engineering challenges \cite{li2023camel, wang2024megaagent, wang2023unleashing}. 

Existing research highlights the potential of LLM agents in simulating collaborative problem-solving environments, where students engage with virtual experts on issues such as ethical dilemmas and environmental trade-offs. These interactions help students develop critical thinking skills and a holistic approach to problem-solving, essential for SDPs. Additionally, MAS and LLMs support adaptive learning by providing real-time feedback, allowing students to experiment and learn from their decisions in a risk-free environment. This active learning approach enhances their ability to navigate complex, globalized engineering challenges \cite{li2023camel, wang2024megaagent}.

\subsection{Existing LLM Multi-Agent Frameworks}

Several LLM-based multi-agent frameworks have emerged, offering advanced capabilities for addressing complex, interdisciplinary challenges. In educational settings, these frameworks are particularly valuable for their ability to simulate real-world scenarios, fostering critical thinking, collaboration, and problem-solving skills. Below are some examples of such frameworks and their key features:

\begin{enumerate}
   
\item Camel-AI \cite{li2023camel} is a communicative agent framework designed to simulate a ``society'' of LLM agents representing different personas. It excels in fostering interdisciplinary dialogue among agents, making it ideal for tasks that require negotiation and the integration of diverse viewpoints. Camel's capacity to explore various facets of ``mind'' interactions makes it a strong candidate for projects where multiple perspectives are needed, such as senior design projects.

\item Crew.AI  \cite{crewAI} is another prominent open-source multi-agent orchestration framework. This Python-based platform allows the orchestration of role-playing AI agents, working together as a cohesive assembly or ``crew'' to complete tasks. The framework's strength lies in automating multi-agent workflows, making it particularly useful in scenarios requiring the coordination of diverse agent roles for collaborative decision-making processes.

\item MegaAgent \cite{wang2024megaagent} is designed to handle large-scale cooperation in MAS. Its primary strength lies in its scalability, enabling a vast number of agents to work together in a coordinated fashion. This makes it suitable for simulating large, complex systems, providing students with insights into how large teams or systems operate in engineering contexts.

\item AgentScope \cite{pan2024very} focuses on multi-agent simulation at a very large scale, providing a highly detailed simulation environment. Its capacity to handle complex, dynamic interactions between agents makes it particularly useful for simulations that require a high degree of realism, such as real-time problem-solving.

\item OpenAgents \cite{xie2023openagents} and Agent Lumos \cite{yin2024agent} are modular frameworks that enable flexible training of LLM agents. OpenAgents provides a platform for building and deploying autonomous language agents in diverse environments, while Agent Lumos unifies and simplifies the training process for these agents. These frameworks are well-suited contexts where the focus is on developing customized agents that can be tuned for specific tasks, such as providing real-time feedback or conducting collaborative discussions.

\end{enumerate}

Together, these frameworks demonstrate the transformative potential of LLM-based multi-agent systems in education. They enable the creation of highly interactive, collaborative environments where students can engage with diverse expert perspectives and gain hands-on experience solving complex engineering problems. These systems provide the flexibility, scalability, and adaptability necessary for real-world problem-solving, making them invaluable tools for enhancing engineering education. For our proposed MAS, we selected Camel AI \cite{li2023camel} due to its use of role-playing and inception prompting, which facilitate agent collaboration with minimal intervention, effectively simulating interdisciplinary teamwork. Additionally, its capability to tackle complex problems by breaking them down into focused, manageable subtasks for each agent makes it an ideal choice.


\section{Methodology}
\label{sec:methodology}

This section presents the methodology used to construct and evaluate our Multi-Agent LLM (MAS LLM) framework for supporting complex engineering problem-solving in SDPs. The methodology is divided into two primary subsections:

\subsection{MAS LLM Framework Construction}
The first subsection outlines the technical details of how the MAS LLM framework was developed. This includes the LLM model used, the integration of the model into a multi-agent system, and the customization of agent roles to simulate diverse expert perspectives such as project managers, breadth and depth agents, and societal and ethical agents. We describe how the system was designed to promote interdisciplinary collaboration and support real-time feedback for students.

Concretely, we designed a copilot-style LLM-powered MAS for engineering and computing students' senior design projects. Fig. \ref{fig:System_Design} illustrates the system design of our proposed MAS. This system comprises eight LLM agents:

\begin{enumerate}
    \item Problem Formulation Agent
    \item Breadth and Depth Agent
    \item Ambiguity and Uncertainty Agent
    \item System Complexity Agent
    \item Technical Innovation and Risk Management Agent
    \item Societal and Ethical Consideration Agent
    \item Methodology and Approach Agent
    \item Comprehensive Evaluation Agent
\end{enumerate}

\begin{figure*}[t]
    \centering
    \includegraphics[width=1\linewidth]{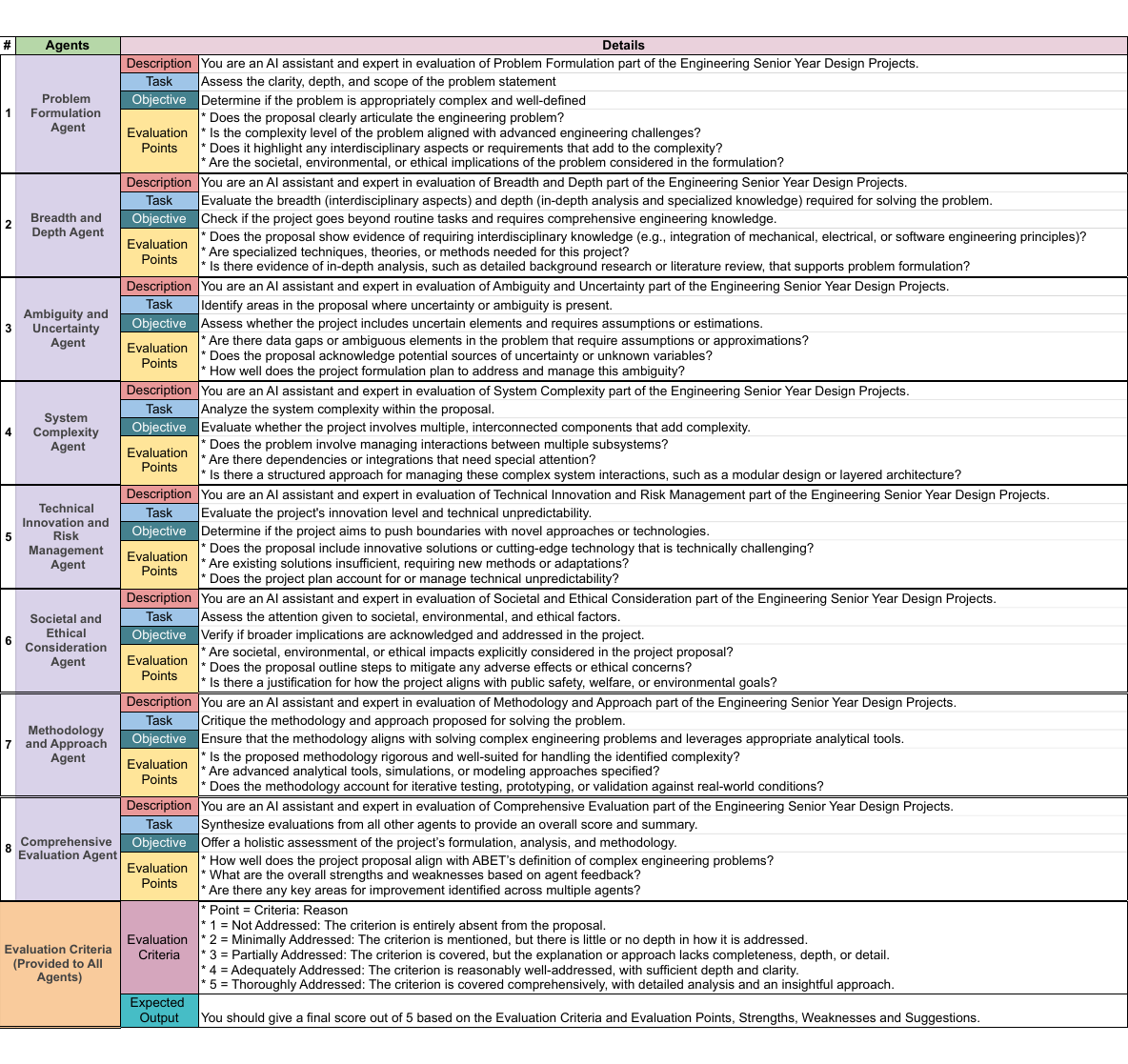}
\caption{Personas for each agent in the proposed MAS, detailing their tasks, objectives, and evaluation points for evaluating engineering SDPs. \textit{The specialized agents—covering problem formulation, breadth and depth, ambiguity, complexity, innovation, ethics, and methodology—enable systematic and holistic assessments across technical and non-technical dimensions.}}
    \label{fig:CompleteAgentsPersonas}
\end{figure*}

Each agent focuses on a different aspect of the SDP based on its role. These agents are designed to act as experts by crafting personas aligned with their respective roles. Box \ref{box:Agent_Persona} shows a sample persona crafted for the ``\textit{Problem Formulation Agent}.'' Each agent receives its persona, project title, and detailed methodology (extracted from PDF) of the proposed SDP (provided by the students). Figure \ref{fig:CompleteAgentsPersonas} shows the detailed personas for each agent used in our proposed MAS. OpenAI's GPT-4o serves as the primary backend LLM for these agents, as it currently outperforms other LLMs \cite{Hello_GPT-4o}. 

The proposed LLM-based MAS operates under a centralized control mechanism managed by a Coordinator Agent, which is responsible for overseeing task execution and ensuring efficient collaboration between agents. Coupled with the Coordinator Agent is a Task Agent, which decomposes each SDP into a series of well-defined tasks for agents to handle. These tasks are then assigned to agents according to their specialized roles through a dedicated Task Channel—a feature in Camel AI designed to streamline task distribution and monitoring \cite{camel_documentation}.

Upon receiving their respective tasks, agents process them sequentially (one by one) and return their outputs to the Task Channel, which then routes them back to the Coordinator Agent. The Coordinator Agent evaluates each agent's output to verify that it meets predefined standards, leveraging a specialized feature provided by Camel AI. As an LLM-based agent, the Coordinator automatically generates a requirements list based on the tasks provided by the user or students. With assistance from the Task Agent, the Coordinator outlines what to expect from the agents, considering the context of the overall task and the sub-tasks assigned to them. If an agent's output does not meet expectations, the Coordinator may reassign the task—either to a different agent with relevant expertise or back to the original agent, augmented with added sub-goals for refinement. When a task from one agent requires input from another agent, the Coordinator efficiently routes the necessary outputs, ensuring both consistency and quality control across the system.

\begin{pabox}[label=box:Agent_Persona]{Problem Formulation Agent Persona}
    \small 
    \textbf{Role:} AI assistant specializing in evaluating the Problem Formulation part of Engineering Senior Year Design Projects.
    
    \textbf{Responsibilities:}
    \begin{itemize}
        \item \textbf{Task:} Assess the clarity, depth, and scope of the problem statement.
        \item \textbf{Objective:} Determine if the problem is appropriately complex and well-defined.
    \end{itemize}
    
    \textbf{Evaluation Points:}
    \begin{itemize}
        \item Does the proposal clearly articulate the engineering problem?
        \item Is the complexity level of the problem aligned with advanced engineering challenges?
        \item Does it highlight any interdisciplinary aspects or requirements that add to the complexity?
        \item Are the societal, environmental, or ethical implications of the problem considered in the formulation?
    \end{itemize}
    
    \textbf{Evaluation Criteria:}
    \begin{itemize}
        \item \textbf{1 = Not Addressed:} The criterion is entirely absent.
        \item \textbf{2 = Minimally Addressed:} The criterion is mentioned, but lacks depth.
        \item \textbf{3 = Partially Addressed:} Criterion covered but lacks completeness.
        \item \textbf{4 = Adequately Addressed:} Criterion is reasonably addressed.
        \item \textbf{5 = Thoroughly Addressed:} Criterion covered comprehensively with insightful analysis.
    \end{itemize}
    
    \textbf{Expected Output:} A final score (1-5) based on Evaluation Criteria and Points, with Strengths, Weaknesses, and Suggestions.
\end{pabox}

Once all tasks are completed, the Coordinator Agent synthesizes the agents' outputs into the final specified format. Each agent operates as a Critic Agent, a Camel AI class designed to provide constructive feedback and facilitate decision-making for complex tasks. Each agent also maintains a message history window of the last 10 responses, enabling it to generate informed outputs. Additionally, agents are equipped with essential tools such as the Internet Search Toolkit and Mathematics Toolkit, provided by Camel AI, to assist in generating accurate and well-rounded responses.  

A snapshot of the user interface for our proposed LLM-based MAS co-pilot is shown in Figure \ref{fig:System_UI}. To enhance interactivity, we provide a follow-up questioning feature where users can ask questions based on the MAS responses. The Coordinator Agent analyzes the follow-up question and previous agent responses to determine which agents are best suited to answer, making the system dynamic and context-aware.

\begin{pabox}[label=box:TOT_Templates]{Tree of Thoughts Prompting Template}
    \small 
`` Imagine X different experts answering this question. \\ 
All experts will write down 1 step of their thinking, and then share it with the group. \\ 
Then all experts will go on to the next step, etc. \\
If any expert realizes they're wrong at any point then they leave. \\
The question is... ''
\end{pabox}

Along with a MAS, we designed a single agent LLM using GPT-4o to evaluate the SDPs on different engineering and computing aspects and to do a systematic and fair evaluation. To make the single agent perform better from its vanilla settings, we utilized the Tree of Thoughts (TOT) prompting technique \cite{hulbert2023using, Saravia_Prompt_Engineering_Guide_2022} to instruct this single agent to simulate the behavior of multiple experts within its response, each covering and evaluating different aspects of the SDP. Box \ref{box:TOT_Templates} shows the template we used to incorporate TOT in our prompting. We used this template as a blueprint for creating a complete prompt for the single agent.


\subsection{Evaluation Methodology} \label{subsec:Evaluation_Methodology}
The second subsection explains how the framework was evaluated in the context of SDPs. We detail the metrics used to assess the effectiveness of the MAS LLM framework in enhancing student learning, collaboration, and problem-solving skills. This includes both qualitative and quantitative data collection methods. We also explain how the data was analyzed to assess the pedagogical and technical impact of the framework.

\subsubsection{Comparing Faculty and Agent Scores}
We designed a methodology to systematically evaluate the performance of this proposed MAS. In a typical SDP progression cycle, students write the initial draft of their proposal and share it with their supervisor/advisor for feedback. Students then incorporate the suggested changes into the proposal, repeating this process multiple times to ensure the proposal meets the standards set by the faculty at their institution. This process takes a significant amount of time for both students and supervisors to prepare the proposal before moving on to the actual development phase of the project. We advocate for the engineering and computing community to adopt this co-pilot system to enhance workflows and facilitate the development of more advanced and effective co-pilot-style MAS for SDP assistance.

\begin{figure*}[t]
    \centering
    \includegraphics[width=\linewidth]{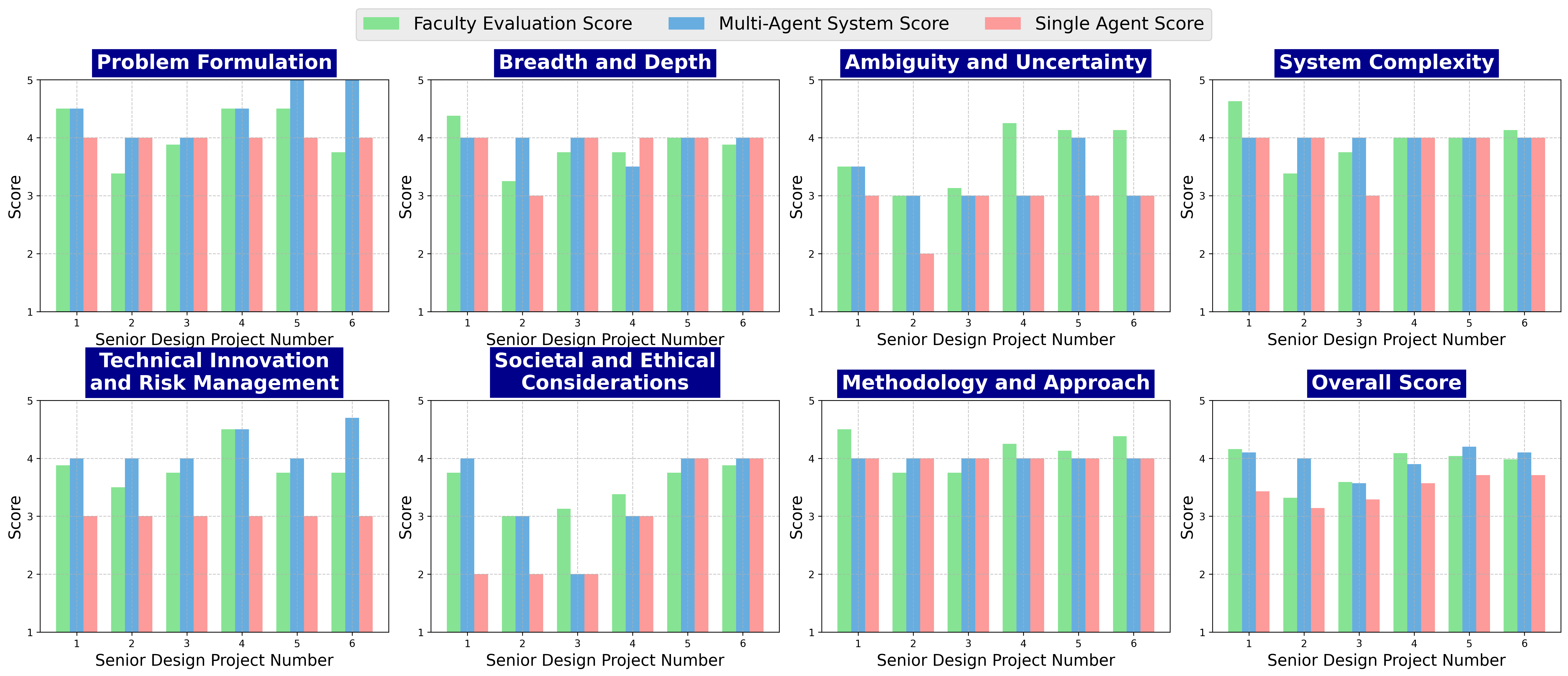}
    \caption{Visual representation of scores to evaluate the performance of both of our proposed multi-agent and single-agent systems across each SDP (represented in numbers on the x-axis) for each aspect of engineering and computing. \textit{It can be seen that multi-agent system scores are more aligned with faculty evaluation scores as compared to single-agent scores.}}
    \label{fig:Engineering_Scores}
\end{figure*}

\begin{figure}
    \centering
    \includegraphics[width=\columnwidth]{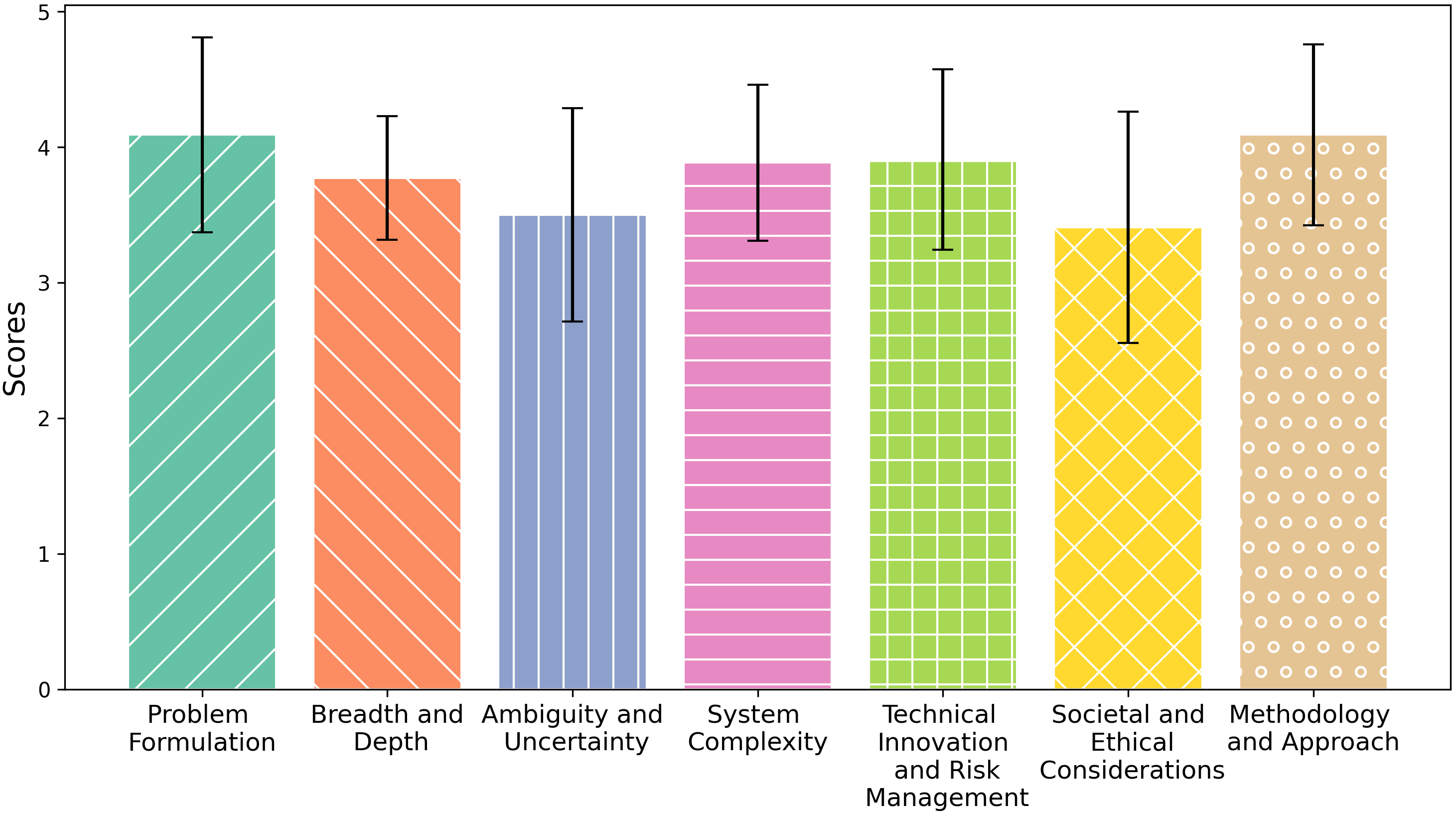}
\caption{Mean Faculty Evaluation Scores with standard deviations for SDP proposals. \textit{The variability illustrates the subjective nature of assessments and the diversity in faculty interpretations of evaluation criteria}.}    \label{fig:FES_STD_MEAN}
\end{figure}

For this evaluation, we asked four faculty members from the Engineering and Computer Science departments at different universities to provide feedback, which we used as a reference standard. We recognize that these evaluations are subjective and naturally vary among evaluators, so they are not an absolute ground truth but serve as a comparative benchmark for our system. The variability in these faculty scores is also shown in Figure \ref{fig:FES_STD_MEAN}. This method uses three primary evaluation scores to assess the effectiveness of our multi-agent system-powered copilot in guiding engineering and computer science students. 
\begin{enumerate}
    \item The \textbf{Faculty Evaluation Score} reflects faculty assessments of key engineering and computing aspects within each SDP proposal, providing a baseline measure of project quality. 
    \item The \textbf{Multi-Agent System Score} represents scores generated by individual agents themselves, each focusing on specific engineering and computing criteria aligned with their programmed expertise. 
    \item The \textbf{Single Agent Score} represents scores generated by a single agent itself using TOT to simulate different experts to evaluate different aspects of the SDP.
\end{enumerate}

Different engineering and computing aspects we selected for this system are Problem Formulation, Breadth and Depth, Ambiguity and Uncertainty, System Complexity, Technical Innovation and Risk Management, Societal and Ethical Considerations, and Methodology and Approach. Together, these scores across each aspect offer a comprehensive view of both the technical quality of student work and the impact of agent-driven feedback in supporting educational outcomes whether it is a MAS or single agent. We collected six SDP proposals from students in the Engineering and Computer Science departments at X University. All students had completed their SDPs (2023-2024). To ensure anonymity, each proposal was renamed with a randomly assigned number (between 1 \& 6). In Section \ref{sec:results}, we will further discuss the evaluation results and performance of our proposed methodology.

\subsubsection{NLP-based Evaluation}
To evaluate the performance of both MAS and single-agent systems, we used four NLP-based scoring metrics: Lexical Cohesion, Average Sentence Length, Clause Density, and Flesch-Kincaid Score. These criteria provide insights into thematic consistency, readability, and structural complexity in system responses.

\begin{enumerate}
     \item  \textbf{Clause Density} \cite{biber2011should} captures sentence complexity by counting clauses per sentence, reflecting layered perspectives. Scores range from 1 (simple, single-idea sentences) to 3+ (highly complex, multi-idea sentences).
     
    \item  \textbf{Lexical Cohesion} \cite{halliday2014cohesion} measures thematic consistency by analyzing word repetition or related terms, indicating how well the content is built on multiple ideas. Scores range from 0 (no cohesion) to 1 (full thematic consistency).

    \item  \textbf{Flesch-Kincaid Score} \cite{kincaid1975derivation} estimates readability, indicating the U.S. grade level needed to understand the text. A higher score suggests advanced content suitable for expert readers, with an ideal range balancing accessibility and sophistication (0–16 scale) for academic purposes.

    \item  \textbf{Average Sentence Length} indicates structural complexity and content depth, with typical ranges from 10 to 40 words. Shorter sentences enhance readability, while longer ones may reflect richer, nuanced perspectives but can be harder to follow. 

\end{enumerate}
These metrics collectively assess the depth and accessibility of each system's response. Using these scores, we can evaluate the performance of these systems from a NLP perspective.


\section{Results}
\label{sec:results}

\begin{figure}
    \centering
    \includegraphics[width=\columnwidth]{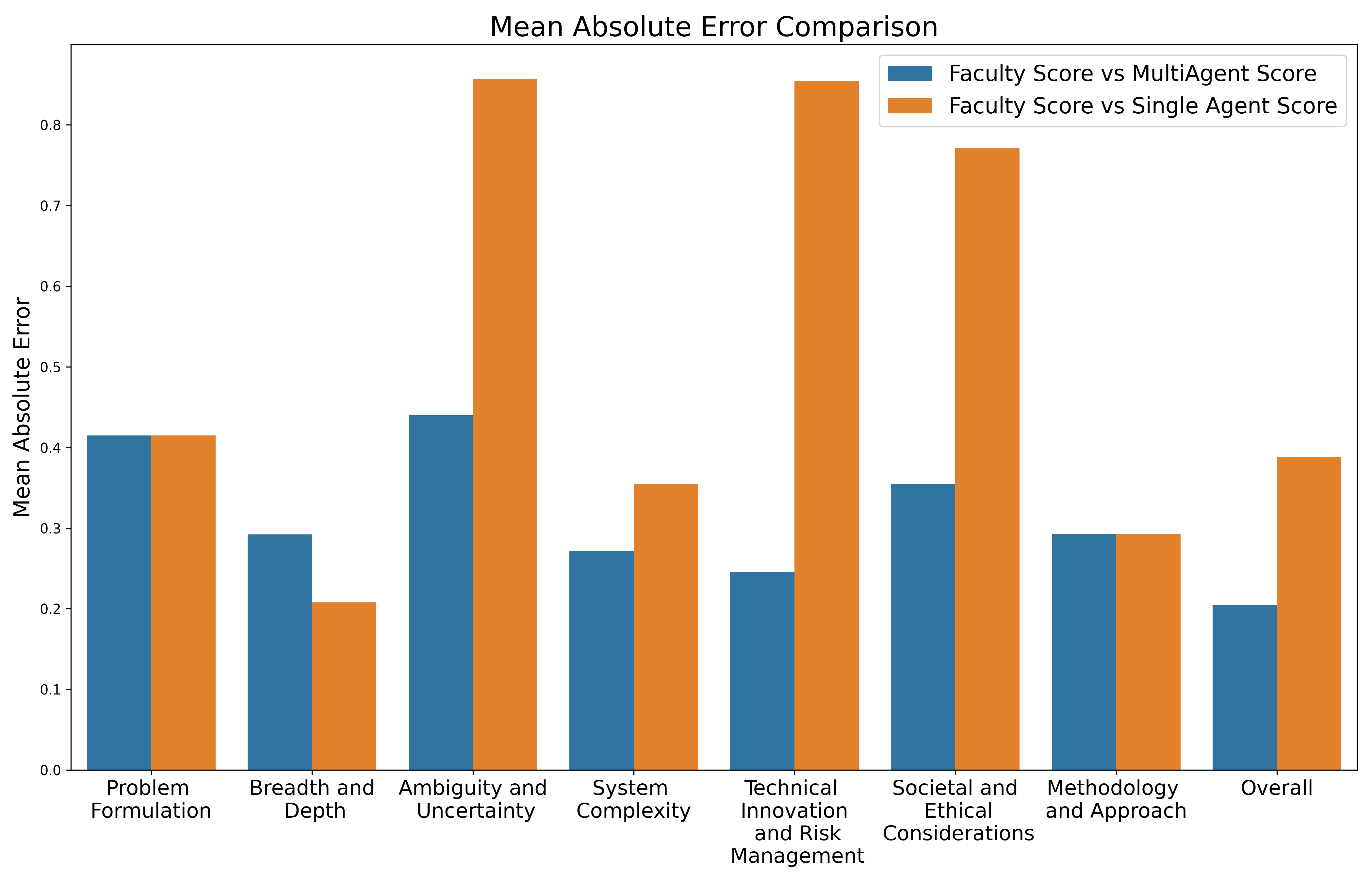}
    \caption{Mean Absolute Error comparison between multi-agent and single-agent systems against faculty evaluations. \textit{Lower bars indicate closer alignment with faculty scores, with the multi-agent system generally showing better accuracy.}}
    \label{fig:MAE}
\end{figure}

\begin{figure*}[t]
    \centering
    \includegraphics[width=1\linewidth]{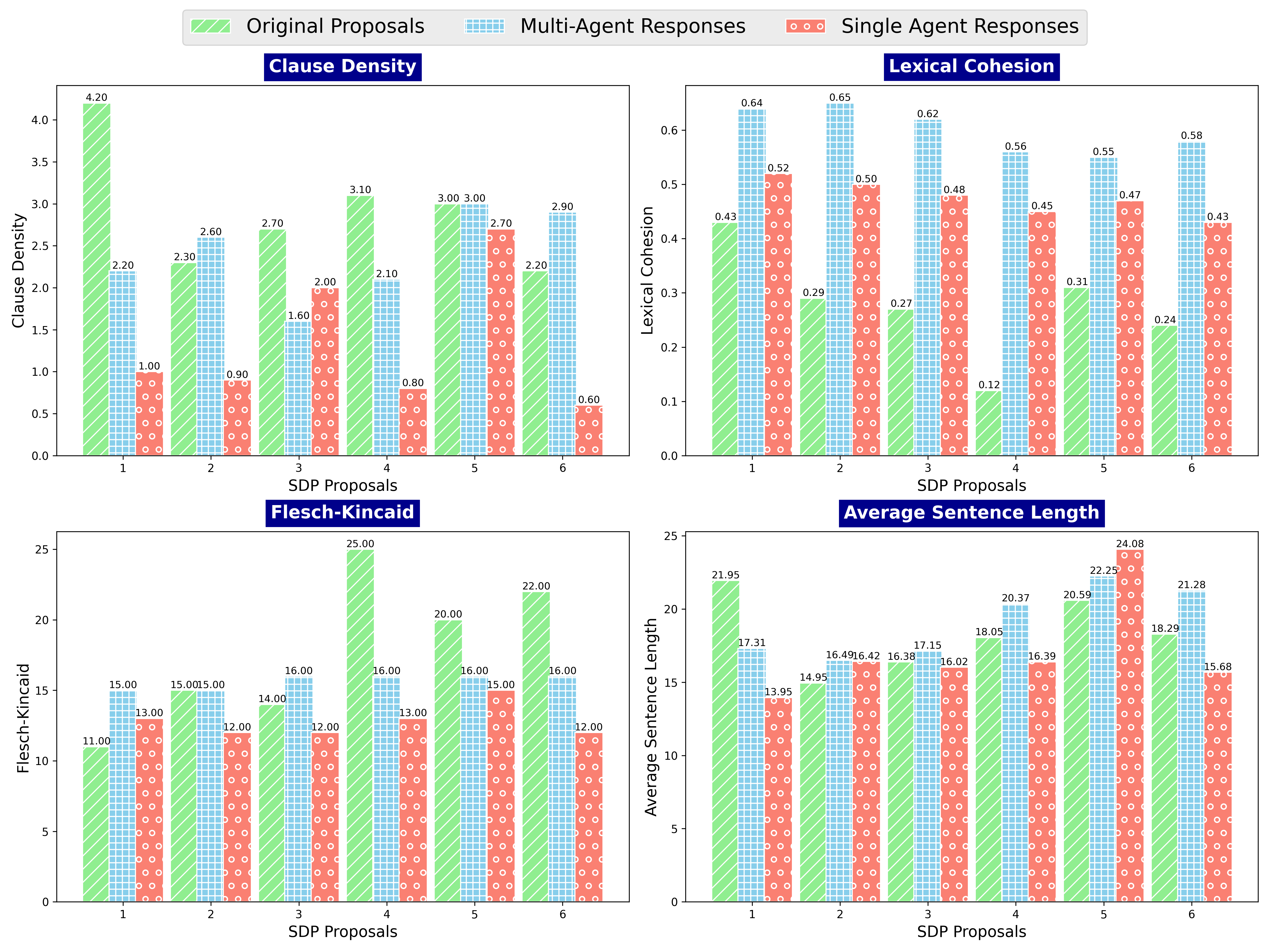}
  \caption{NLP-based performance evaluation of original student proposals, MAS responses, and single-agent system responses. The metrics include Clause Density (complexity), Lexical Cohesion (thematic unity), Flesch-Kincaid Score (readability), and Average Sentence Length (structural depth), designed to balance accessibility with scholarly depth. \textit{Our results show that the MAS approach consistently outperforms the single-agent system across all evaluated metrics}.}
   \label{fig:combined_nlp_metrics}
\end{figure*}

In this section, a comprehensive analysis of the results is presented, focusing on faculty evaluations, error rates, and NLP-based performance metrics. The evaluation based on faculty scores is detailed in Figure \ref{fig:Engineering_Scores}. Error rates for each system, reflecting their relative accuracy, are shown in Figure \ref{fig:MAE}. Finally, the performance of each system, assessed using NLP-based metrics, is illustrated in Figure \ref{fig:combined_nlp_metrics}.

\subsection{Faculty Guided Evaluation Results} \label{subsec:Faculty_Guided_Eval}
As outlined in Section \ref{subsec:Evaluation_Methodology}, we designed three distinct evaluation scores to assess the performance of our proposed MAS for SDPs with respect to evaluation from faculty members. Scores comparison for each project across faculty evaluations, MAS scores, and single-agent scores. Green bars represent faculty scores, blue bars represent MAS scores and red bars indicate single-agent scores. This section further discusses key insights from the observed scoring patterns and analysis.

The detailed results in Figure \ref{fig:Engineering_Scores} show the performance of each system across each aspect and SDP proposals. The graph shows that the MAS consistently matches or outperforms the single-agent system across all aspects, except for an isolated project for the Breadth and Depth aspect. This can be seen by the MAS score in each aspect being much closer to the faculty evaluation scores and following the same trend. 

Figure \ref{fig:MAE} presents the results showcasing the effectiveness of the MAS in evaluating SDPs compared to the single-agent system, with both systems benchmarked against faculty evaluation scores. The MAS demonstrates greater alignment with faculty evaluations, with a Mean Absolute Error (MAE) of 0.205 compared to 0.388 for the single-agent system, an 89.3\% accuracy improvement. Lower bars in the figure represent closer alignment with faculty scores. The MAS excels in technical categories such as Technical Innovation and Risk Management (MAE 0.345 vs. 0.855) and System Complexity (MAE 0.272 vs. 0.355). However, in Breadth and Depth, the single-agent system performs better (MAE 0.208 vs. 0.292), suggesting certain holistic aspects may favor a unified approach. The MAS outperforms in Ambiguity and Uncertainty (MAE 0.440 vs. 0.857) and Societal and Ethical Considerations (MAE 0.355 vs. 0.772), demonstrating its broader effectiveness. Both systems perform equally in Methodology and Approach (MAE 0.293).

These results demonstrate the effectiveness of the MAS-based approach in evaluating SDPs, with superior accuracy across most assessment criteria due to its use of specialized agents. While the MAS consistently excels in key areas, including technical depth and ethical considerations, it lags behind the single-agent system in the Breadth and Depth category. This highlights an area for further refinement, reinforcing the MAS's potential as a robust and reliable tool for academic project evaluation with targeted improvements.

\subsection{NLP-based Evaluation Results} \label{subsec:NLP_based_Eval}
From an NLP perspective, we designed a mechanism to evaluate the responses of the MAS and single-agent systems. These metrics assess how each system performed in terms of complex ideation, structural coherence, and readability for both students and supervisors.

As shown in Figure \ref{fig:combined_nlp_metrics}, the clause density of responses generated by the MAS is more closely aligned with the original proposals compared to those produced by the single-agent system. This indicates that the MAS outputs are more detail-rich, conveying a greater amount of information per sentence, which enhances their effectiveness and suitability in this context. Examining the lexical cohesion graph in Figure \ref{fig:combined_nlp_metrics}, we see that MAS responses exhibit stronger thematic consistency. Unlike the original proposals written by students and the single-agent responses, MAS outputs offer more collaborative feedback reflecting the interconnectedness of ideas within the text. With different agents covering specific aspects and building upon each other's responses, MAS produces nuanced, detailed outputs that better support both students and supervisors.

The Flesch-Kincaid readability score, which indicates the U.S. grade level needed to comprehend the responses on a first read, is a crucial metric for evaluating academic suitability. As grade level increases, writing typically becomes more organized and detail-rich---a pattern common in academic documents. For senior-year students engaged in final-year SDP projects, an ideal readability score lies between 14 and 16. As shown in Figure \ref{fig:combined_nlp_metrics}, the MAS responses in the Flesch-Kincaid graph, predominantly fall within this ideal range, suggesting they are well-suited to the academic level of senior students. The Average Sentence Length metric in this figure indicates that sentence lengths in both MAS and SA responses are similar to those in the original proposals in most of the cases. While typical values for this metric range from 0 to 40, responses from both the MAS and the single-agent system are predominantly concentrated in the mid-range, often aligning more closely with the original proposals. This effect is primarily attributed to the training methodology of the LLMs. Given that both the MAS and single-agent systems rely on the same backend LLM, this behavior is unsurprising.

The results in both faculty-based and NLP-based evaluations (\S\ref{subsec:Faculty_Guided_Eval} and \S\ref{subsec:NLP_based_Eval}) highlight the superior performance characteristics of the MAS compared to the single-agent system in evaluating SDPs, with each system displaying distinct strengths. MAS responses demonstrated broader coverage, not only in technical aspects, clause density, and thematic consistency but also in providing more detailed, cohesive feedback. Readability scores indicate that MAS responses align well with senior-year academic expectations, although slightly longer sentence lengths may impact accessibility. Additionally, the MAS exhibited impressive performance in addressing ethical and societal considerations. The single-agent system, on the other hand, displayed higher error rates across most metrics, except in areas like Breadth and Depth, where its error rate is lower than MAS, aligning it closely with faculty scores. This systematic evaluation concretely shows that MAS-based LLMs are more suited for complex problem-solving in engineering and computer science fields because of the multifaceted nature of the requirements for SDPs.

\section{Discussions}
\label{sec:discussions}

\begin{figure*}[!t]
    \centering
    \includegraphics[width=0.9\linewidth]{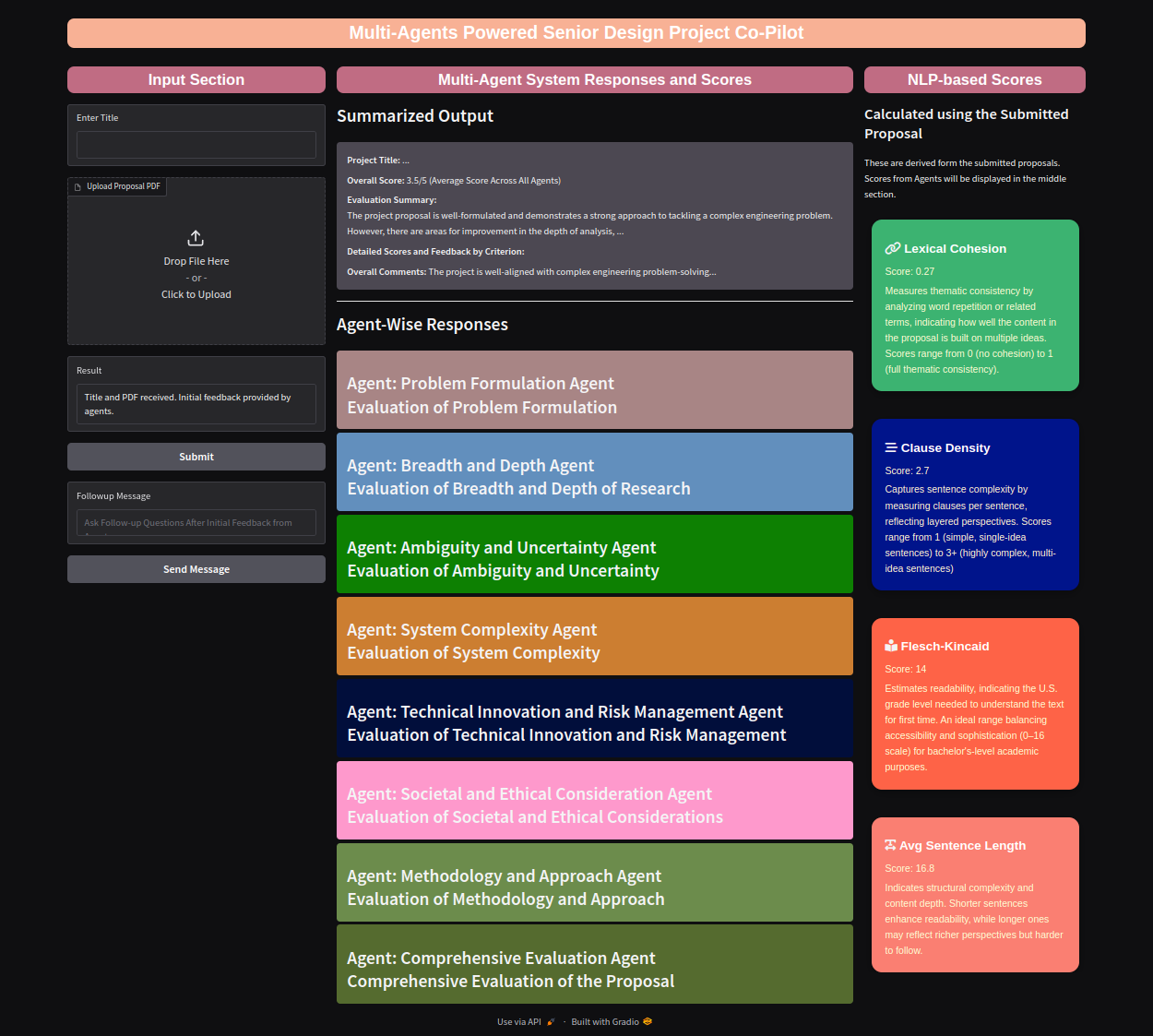}
\caption{User interface for the proposed LLM-based MAS co-pilot, enabling students to interact with a graphical interface. Students can upload their SDP title and proposal, view summaries and agent feedback, and submit follow-up questions via the interface. \textit{This interface streamlines the feedback process, promoting iterative learning and comprehensive project assessment.}}    \label{fig:System_UI}
\end{figure*}

\subsection{Situating our Initial Approach in the Agentic Landscape}

The landscape of LLM agents and MAS is diverse, with various frameworks actively developed to leverage large language models for complex, collaborative tasks. MAS, a longstanding area of research since the 1980s, provides a foundation for decentralized, autonomous problem-solving across domains like artificial intelligence and robotics \cite{wooldridge2009introduction}. MAS frameworks facilitate agent interactions through mechanisms such as cooperation, coordination, and negotiation, often yielding emergent behaviors beyond the capabilities of individual agents. 

While powerful, LLMs alone lack the sophistication to fully function as autonomous agents, as they typically need plugins or external tools to interact meaningfully with other systems. In MAS, agents are traditionally defined as systems capable of autonomous action and interaction within an environment to achieve specific goals \cite{wooldridge2009introduction}. MAS frameworks allow agents to coordinate, negotiate, and cooperate, often resulting in emergent behaviors beyond the reach of individual agents. 

In GenAI, however, an agent is defined as a GenAI system, usually powered by an LLM, that serves a user's goals by engaging with external systems and executing actions outside the LLM itself \cite{schulhoff2024prompt}. For example, ChatGPT's code interpreter, integrated into GPT-4, combines language processing with Python to autonomously handle tasks like data analysis and plotting based on user prompts, improving accuracy by performing direct calculations. Yet, this setup still requires human oversight and does not fully align with the traditional concept of an agent. 

There is now rising interest in combining the power of agents and LLMs\footnote{\url{https://llmagents-learning.org/f24}}. Advanced LLM Agent frameworks, such as the MRKL System \cite{karpas2022mrkl}, extend LLMs into more sophisticated roles by employing a central LLM router to access various tools for complex problem-solving. Further enhancing agent autonomy, systems like LangChain \cite{topsakal2023creating}, AutoGen \cite{wu2023autogen}, and ReAct \cite{yao2022react} minimize continuous human input, and Camel AI \cite{li2023camel} leverages role-playing and inception prompting to enable agents to collaborate with minimal intervention, simulating interdisciplinary teamwork. Building on the established use of role-playing in engineering education as a pedagogical tool \cite{hingle2024framework}, and the importance of formative feedback and assessment \cite{qadir2020leveraging}, we extend this concept by harnessing the capabilities of multi-agent LLMs.

In this work, we introduce a Camel AI-inspired framework for SDPs that provides a structured, role-based problem-solving environment, simulating interdisciplinary collaboration to help students tackle complex engineering tasks. Our system marks an initial exploration into LLM-based multi-agent systems, and unlike more advanced frameworks, it lacks advanced external tool integration and sophisticated multi-agent techniques like coordination and negotiation, which could further enhance collaboration. Despite not having all the advanced techniques our proposed MAS still had 89\% more accuracy compared to single-agent systems. Emerging methods for optimizing LLM agent interactions, including advanced prompting techniques \cite{promptingguide2024}, are also beyond our current scope. Future work will explore these areas, to develop a more autonomous multi-agent system that leverages complex coordination and external tools to meet the demands of increasingly sophisticated engineering challenges.

\subsection{Pedagogical Implications}

Our work impacts key educational stakeholders: students, instructors, and administrators. For students, the multi-agent LLM system provides structured guidance through a web-based co-pilot, helping them tackle complex, interdisciplinary projects while fostering critical thinking without requiring full subject expertise.  Figure~\ref{fig:System_UI} illustrates a snapshot of the user interface for the proposed MAS co-pilot system.  Given that students are often engaging with complex problem-solving for the first time, our framework offers a practical tool that helps bridge their knowledge gaps and guides them in areas like ethical considerations, technical complexity, and societal impact. However, as current LLMs still rely heavily on human input and precise prompt engineering, students who lack experience in both prompt design and advanced subject nuances may face challenges. This system addresses some of these gaps by offering structured pathways for problem-solving, thus enhancing their ability to work on large-scale projects. For instructors, the framework improves instructional efficiency by automating guidance in key project areas, allowing them to focus on high-level mentorship rather than technical troubleshooting. For administrators, the system can support accreditation efforts by providing a scalable tool that aligns with goals for multidisciplinary and real-world education. 

To encourage further development and adaptation, we have shared the code for our SDP complex problem-solving co-pilot as an open-source resource \textit{(Link will be released after paper acceptance)}, allowing educators and researchers to experiment with and extend the system. By making our framework accessible, we aim to contribute to the wider adoption of advanced multi-agent LLM systems in engineering education, fostering a collaborative movement toward practical tools that support complex problem-solving in educational contexts.

\section{Conclusions}
\label{sec:conclusions}

In this paper, we explored the use of Multi-Agent Large Language Models (MAS LLMs) as a novel framework for enhancing complex problem-solving in engineering senior design projects. By leveraging the collaborative capabilities of MAS and the generative power of LLMs, we provided students with a dynamic, interdisciplinary environment that mirrors real-world engineering challenges. This approach enabled students to engage with multiple perspectives, simulate trade-offs, and explore the complexity of decision-making in globalized engineering contexts. We evaluated our MAS LLMs and a single-agent system based on some custom-designed metrics that are more inclined toward faculty members of universities and some widely adopted NLP-based metrics. Our findings suggest that the MAS LLM framework can enhance student learning by providing details-rich responses, ideation of complex ideas, fostering collaboration, improving problem-solving skills, and allowing for real-time feedback through follow-up questions. With MAS achieving an overall accuracy of 89\%. Our analysis also reveals that single-agent systems are not usually aligned with faculty-assigned scores in most of the engineering and computing aspects. This is due to the fact that the single-agent system tries to oversimplify the responses, and the response window of LLMs is much smaller for one agent as compared to multiple responses from each agent in MAS. The ability of MAS to simulate expert personas, representing diverse stakeholders and viewpoints, and responding with higher attention to details and complexity of senior design projects, offers students a deeper understanding of the ethical, technical, and social dimensions of their projects. Moreover, the framework's adaptability to various project types and disciplines makes it a promising tool to be used and evaluated by the engineering and computing education community. While our study demonstrates the potential of MAS LLMs in educational settings, future work could explore further customizations of agent behavior and extend the framework to other fields beyond engineering and computer science. Additionally, the integration of more advanced evaluation metrics and longitudinal studies could provide deeper insights into the long-term educational benefits of this approach.

\bibliographystyle{IEEEtran}
\bibliography{refs}

\end{document}